\documentclass[aps,amsmath,amssymb,reprint,superscriptaddress]{revtex4-1}
\usepackage[T1]{fontenc}
\usepackage{mathptmx}
\usepackage{graphicx}
\usepackage[colorlinks=true, citecolor=blue, urlcolor=magenta, linkcolor=blue, anchorcolor=red]{hyperref}

\begin{document}

\title{Longitudinal elastic wave control by pre-deforming semi-linear materials}

\author{Dengke Guo}
\thanks{These two authors contributed equally}
\affiliation{School of Aerospace Engineering, Beijing Institute of Technology, 100081, Beijing, China}

\author{Yi Chen}
\thanks{These two authors contributed equally}
\affiliation{School of Aerospace Engineering, Beijing Institute of Technology, 100081, Beijing, China}

\author{Zheng Chang}
\affiliation{College of Science, China Agricultural University, Beijing 100083, China}

\author{Gengkai Hu}
\email{hugeng@bit.edu.cn}
\affiliation{School of Aerospace Engineering, Beijing Institute of Technology, 100081, Beijing, China}

\begin{abstract}
An incremental wave superimposed on a pre-deformed hyper-elastic material perceives an elastic media with the instantaneous modulus of the current material. This offers a new route with a broadband feature to control elastic waves by purposely creating finite deformation field. This study proves that the governing equation of a semi-linear material under a symmetric pre-deformation condition maintains the form invariance for longitudinal wave, so the longitudinal wave control can be made by transformation method without the constraint condition on principle stretches, but this is not the case for shear waves. Therefore pre-deforming a semi-linear material provides a potential method for treating longitudinal and shear waves differently. Examples with elastic wave control and band structure shift through pre-deforming a semi-linear material are provided to illustrate this finding. Finally, a one-dimensional spring lattice is proposed to mimic a semi-linear material, and the dispersion relation for longitudinal waves in a sandwich structure with such spring lattice is shown to be invariant during elongation, confirming the result found based on a homogeneous semi-linear material. These results may stimulate researches on designing new hyper-elastic microstructures as well as designing new devices based on pre-deformed hyper-elastic materials.
\end{abstract}
\maketitle

\section{INTRODUCTION}
Wave steering by carefully distributing material in space is an interesting topic for both scientific and engineering communities. A recent significant progress along this line is the proposition of transformation method\cite{pendry2006controlling,milton2006cloaking,cummer2007one,norris2008acoustic}, which in essence maps physical fields from one region to another based on the form-invariance of the governing equation. Within the framework of conventional elasticity, the elastic wave equation cannot retain its form under a general mapping\cite{milton2006cloaking,norris2011elastic}. An approximate method\cite{chang2010design} with conventional elastic materials or an exact control strategy\cite{brun2009achieving} with hypothesized elastic materials without the elastic minor symmetry are proposed to control elastic waves. Another interesting method is to look at the degenerated elastic materials, for example, pentamode materials are shown to satisfy the form-invariance\cite{norris2008acoustic}, but they are used to control specifically the pseudo pressure wave instead of a fully coupled elastic wave. The transformation method based on pentamode materials generally requires a symmetric or quasi-symmetric mapping\cite{chen2016design}. Realization of the materials to meet the design requirement is of a great challenge and needs meticulous microstructure design\cite{buckmann2014elasto,chen2017broadband}. In addition, they may suffer from limited frequency band due to inevitable dispersion at high frequency.

To overcome these shortcomings, an alternative promising route is proposed to steer elastic waves by pre-deforming a hyper-elastic material. Wave on the pre-deformed material perceives a new media with the modified local density and instantaneous modulus of the current material\cite{ogden2007incremental}. This effect purely originates from the finite deformation and hyper-elastic constitutive relation of the hyper-elastic material. The non-dispersive local density and instantaneous modulus may provide a new way to control elastic waves with a broadband efficiency. By noticing the similarity between the material parameters required by the asymmetric transformation\cite{brun2009achieving} and the local density and instantaneous modulus\cite{ogden2007incremental} of a pre-deformed hyper-elastic material, the hyper-elastic transformation theory\cite{parnell2012nonlinear,norris2012hyperelastic,chang2015disentangling} is proposed to control the superimposed incremental waves by pre-deforming a hyper-elastic material. Neo-Hookean1\cite{parnell2012nonlinear,parnell2012employing} hyper-elastic materials have been investigated to design a cylindrical cloak for anti-plane shear waves, the cloaking effect is demonstrated from scattering analysis. This hyper-elastic material is also used to design a phononic crystal with invariant band gaps\cite{barnwell2016antiplane} for shear waves. Chang et al.\cite{chang2015disentangling} further studied the in-plane longitudinal and shear waves in a compressible neo-Hookean material under a pre-shear deformation, and they found that the in-plane shear waves will follow the design by the hyper-elastic transformation while the longitudinal waves will not be affected. However, it is only rigorously proved that incompressible neo-Hookean materials maintain the form invariance for anti-plane shear waves\cite{parnell2012nonlinear,parnell2012employing,barnwell2016antiplane}. Norris and Parnell proposed to use a semi-linear hyper-elastic material with a symmetric pre-deformation to control fully coupled elastic waves\cite{norris2012hyperelastic}, and a cylindrical cloak is designed for both in-plane longitudinal and shear waves. Based on the semi-linear energy function in terms of three principal stretches, they find that the out of plane extensions should be adapted accordingly with the in-plane stretches in order to control the in-plane longitudinal and shear waves\cite{norris2012hyperelastic}, simultaneously. In this paper, we observe that the second order derivatives of the semi-linear energy function are composed of the deformation gradients and a fourth order tensor with the minor anti-symmetry. This forth order tensor with the minor anti-symmetry is cancelled out with a divergence field, implying that a semi-linear material, under a general symmetric deformation without the out of plane extensions constraint, still maintains the form invariance for a divergence field. Therefore the transformation method can be used to control longitudinal waves. This finding provides a useful tool to design devices interacting differently with the longitudinal and shear components of a fully coupled elastic wave.

This research is arranged as follows. First, we will prove that the divergence component of elastic waves on a symmetrically pre-deformed semi-linear material can be strictly mapped from the initial configuration. These results indicate that, the governing equation for longitudinal waves in a semi-linear material under a symmetric pre-deformation maintain the form invariance, so they can be controlled within the strategy of transformation method by pre-deformation. Second, numerical simulations, demonstrating wavelength magnifying and dispersion relation invariance for longitudinal waves on a pre-deformed semi-linear material, will be conducted to validate the transformation method. Finally, a spring lattice is proposed as a one-dimensional (1D) prototype for a homogeneous semi-linear material, the invariant dispersion relation for longitudinal waves under a finite elongation will be demonstrated through a 1D sandwich structure to confirm the theoretical finding.

\section{METHODS}
\subsection{Wave propagation on a deformed hyper-elastic material}

A stress free hyper-elastic material occupying an initial configuration $\Omega_{0}$ is deformed to the current configuration $\Omega$ through the boundary and body loads. Mathematically, this process is represented by a mapping between the two configurations $\mathbf{x}:\Omega_{0}\rightarrow\Omega$, which maps any point $\mathbf{X}$ in $\Omega_{0}$ to another point $\mathbf{x}$ in $\Omega$, $\mathbf{x}=\mathbf{x}(\mathbf{X})$. This mapping can also be denoted by the deformation gradient $\mathbf{F}=\partial\mathbf{x}/\partial\mathbf{X}$ in nonlinear elasticity\cite{ogden2007incremental}, or in component form $F_{iJ}=\partial x_{i}/\partial X_{J}$. Here, lowercase and uppercase subscripts refer to the initial configuration and current configuration, respectively. The deformation tensor $\mathbf{F}$ in general can take any form compatible with physical deformation, e.g., identity, diagonal or symmetric form.
A small displacement perturbation $\mathbf{u}(\mathbf{x}, t)$ superimposed on the deformed body $\Omega$ is governed by the dynamic equation in the current configuration\cite{ogden2007incremental},
\begin{equation}
\label{eq1}
\rho u_{j,tt}=\nabla_{i}(C_{ijkl}\nabla_{k}u_{l})
\end{equation}

Here, $i$ with a lowercase subscript representing the partial derivative with respect to the current coordinate $\mathbf{x}_i$, the displacements ul with a lowercase subscript mean the fields physically occurred in the current configuration, and subscript $t$ denotes time derivative. The used subscripts range from 1 to 3 and duplicated indexes should be understood as summation. The local density $\rho$ and instantaneous elastic tensor $\mathbf{C}$ in the current configuration are,
\begin{subequations}
\begin{equation}
\rho=J^{-1}\rho_0 \label{eq2a}
\end{equation}
\begin{equation}	C_{ijkl}=J^{-1}F_{iM}F_{kN}A_{MjNl}=J^{-1}F_{iM}F_{kN}\frac{\partial^2W}{\partial F_{jM} \partial F_{lN}} \label{eq2b}
\end{equation}
\end{subequations}
In which, $J$ is determinant of the deformation tensor $\mathbf{F}$,  $\rho$ is derived from the mass conservation law and  $\rho_0$ represents the density in the initial configuration, $W$ is the strain energy function of the hyper-elastic material. The instantaneous elastic tensor $\mathbf{C}$ is a forth order tensor without the minor symmetry, and the incremental wave is likely to be governed by an elastic material with asymmetric incremental stress  $C_{ijkl}\nabla_{k}u_l$. The true Cauchy stress in the hyper-elastic material is indeed still symmetric, only the incremental stress is asymmetric. This observation stimulates researches on the transformation theory for elastic waves through a finite deformation in a hyper-elastic material\cite{parnell2012nonlinear,norris2012hyperelastic,chang2015disentangling}. It should be noted, although the actual perturbation occurs on the current configuration $\Omega$, its governing equation can be transformed to the initial configuration $\Omega_{0}$. Following the finite deformation elasticity theory\cite{ogden2007incremental}, Eq. (\ref{eq1}) can be easily pulled back to the initial configuration,
\begin{equation}
\label{eq3}
\rho_0 u_{j,tt}=J\nabla_{i}(J^{-1}F_{iM}F_{kN}A_{MjNl}\nabla_{k}u_{l})=\nabla_M(A_{MjNl}\nabla_N u_l)
\end{equation}
In above equation, two mathematic identities, $\nabla_i(J^{-1}F_{iM})=0$, $F_{iM}\nabla_i=\nabla_M$ are used. $\nabla_M$ with an uppercase subscript represents the partial derivative with respect to the initial coordinate $\mathbf{X}_M$. It is seen that the modulus, $A_{MjNl}=\partial^2 W/\partial F_{jM}/\partial F_{lN}$, functions as a nominal elastic modulus in the initial configuration, and therefore is called the pull-backed modulus in the following.

For the same hyper-elastic material without the pre-deformation, wave propagation is governed by the equation in the initial configuration $\Omega_{0}$,
\begin{equation}
\begin{split}
\label{eq4}
\rho_0 v_{J,tt}=&\nabla_M(C_{0mjnl}\nabla_N v_l) \\
=&(\lambda_0+\mu_0)\nabla_J(\nabla_L v_L)+\mu_0\nabla_M\nabla_M v_J
\end{split}
\end{equation}
Here, $\mathbf{v}(\mathbf{X}, t)$ is used for the displacement fields on the hyper-elastic material without the pre-deformation to distinguish from the displacement field $\mathbf{u}(\mathbf{x}, t)$ with the pre-deformation. The displacements $v_J$ with an uppercase subscript mean the fields physically occurred in the initial configuration. Apparently, the above equation is a special case of Eq. (\ref{eq3}) with the deformation tensor being an identity matrix $\mathbf{F}=\mathbf{I}$, and the elastic tensor for a conventional isotropic elastic material is recovered, i.e., $C_{0mjnl}=\lambda_0\delta_{mj}\delta_{nl}+\mu_0(\delta_{mn}\delta_{jl}+\delta_{ml}\delta_{jn})$ with $\lambda_0$ and $\mu_0$ being the lam\'e constants.

\subsection{Transformation method for longitudinal waves with semi-linear materials}

In the following, we will prove the governing equation for longitudinal waves is form-invariant for semi-linear materials under a symmetric deformation. The strain energy function of a semi-linear material writes\cite{norris2012hyperelastic},
\begin{equation}
W=\frac{\lambda_0}{2}(U_{KK}-3)^2+\mu_0(U_{KL}-\delta_{KL})(U_{KL}-\delta_{KL})
\end{equation}
In which, $\mathbf{U}=\mathbf{R}^{-1}\mathbf{F}$ is the right stretch tensor for the finite deformation gradient $\mathbf{F}$ and $\mathbf{R}$ is an orthogonal matrix $\mathbf{R}^{\mathbf{T}}\mathbf{R}=\mathbf{I}$. The strain energy function $W$ is only related to the stretch tensor $\mathbf{U}$ as required by objectivity. The pull-backed modulus $A_{MjNl}$ for a general deformation gradient $\mathbf{F}$ is given in \textbf{Appendix}\ref{appen}, and can be further simplified when $\mathbf{F}$ is symmetric with $\mathbf{R}=\mathbf{I}$,
\begin{equation}
\label{eq6}
\begin{split}
A_{MjNl}|_{\mathbf{F}=\mathbf{F}^{\mathbf{T}}}=&\lambda_0\delta_{jM}\delta_{lN}+2\mu_0\delta_{ij}\delta_{MN}\\
+&(\lambda_0(U_{KK}-\delta_{KK})-2\mu_0)\frac{\partial R_{jM}}{\partial F_{lN}}
\end{split}
\end{equation}
In addition to the first two constant terms, the pull-backed modulus has a third term depending on the deformation gradient $\mathbf{F}$. From the geometric point of view, the symmetry of the deformation tensor $\mathbf{F}$ means the deformation field is irrotational. If $\mathbf{F}$ further becomes an identity matrix $\mathbf{I}$, the modulus $A_{MjNl}$ will degenerate to the conventional isotropic elasticity tensor (see more details in \textbf{Appendix}\ref{appen}). Substituting the above pull-backed modulus Eq. (\ref{eq6}) into Eq. (\ref{eq3}) and taking the divergence operation on both sides, one will obtain the following equation,
\begin{equation}
\begin{split}
\rho\nabla_J u_{j,tt}&=\nabla_J\nabla_M(A_{MjNl}\nabla_N u_l)\\
&=(\lambda_0+2\mu_0)\nabla_M\nabla_M(\nabla_L u_l)
\end{split}
\end{equation}
In deriving the above equation, the skew-symmetry of $\partial R_{jM}/\partial F_{lN}$ with respect to index $j$ and $M$ is used. This equation is exactly the same as the governing equation for the divergence field on the initially un-deformed configuration, i.e., $\rho_0\nabla_J v_{J,tt}=(\lambda_0+2\mu_0)\nabla_M\nabla_M(\nabla_J v_J)$. In this case, once the boundary and loading conditions for the two configurations are set accordingly\cite{chen2016design}, the divergence field on the pre-deformed body can be mapped from that on the initial configuration $\nabla_L u_l=F_{kJ}(\nabla_k u_j)=\nabla_J v_J$. This observation implies that, if one region of a semi-linear material is pre-deformed without altering its boundary, then an impinged longitudinal wave inside the pre-deformed region will follow the mapping, while outside the pre-deformed region it will not be influenced at all and just perceives a homogeneous semi-linear material. It can be proved similarly that the curl field on the pre-deformed semi-linear material doesn't follow the same equation as on the initial configuration, and therefore cannot be mapped. This property can be used to design wave controlling devices interacting differently with longitudinal and shear waves.

A more explicit expression for $A_{MjNl}$ is not available for a general symmetric gradient, however, a very simple formula for the in-plane components is possible for the special symmetric gradients with $F_{13}=F_{23}=0$, $F_{33}\not=0$,
\begin{equation}
\label{eq8}
\begin{split}
A_{MjNl}|_{\mathbf{F}=\mathbf{F}^{\mathbf{T}}}=&\lambda_0\delta_{jM}\delta_{lN}+2\mu_0\delta_{ij}\delta_{MN}\\
+&(\lambda_0(U_{kk}+U_{33}-3)-2\mu_0)\frac{1}{F_{kk}}\epsilon_{jM}\epsilon_{lN}
\end{split}
\end{equation}
Here, all the index $M$, $j$, $N$, $l$ and $k$ range from 1 to 2, $\epsilon_{jM}$ is the 2D permutation tensor. Further if the out of plane stretch is constraint accordingly with the in-plane stretches,
\begin{equation}
\label{eq9}
U_{33}=1-\frac{\lambda_0+\mu_0}{\lambda_0}(U_{kk}-2)
\end{equation}
Equation (\ref{eq9}) is the same as (4.14) by Norris and Parnell\cite{norris2012hyperelastic}. Substituting Eq. (\ref{eq9}) into Eq. (\ref{eq8}), and noticing $U_{ηη}=F_{ηη}$, one will obtain an isotropic in-plane elastic tensor,
\begin{equation}
\label{eq10}
A_{MjNl}|_{\mathbf{F}=\mathbf{F}^{\mathbf{T}}}=\lambda_0\delta_{jM}\delta_{lN}+\mu_0(\delta_{ij}\delta_{MN}+\delta_{jN}\delta_{Ml})
\end{equation}
So with the constraint condition of Eq. (\ref{eq9}), both longitudinal and in-plane shear waves can be controlled simultaneously with transformation method, as observed by Norris and Parnell\cite{norris2012hyperelastic}. However we demonstrate here, for longitudinal waves the constraint condition can be removed.

\section{NUMERICAL EXAMPLES}
\subsection{Wave propagation on a symmetrically pre-deformed semi-linear material}

Numerical simulation based on the hyper-elastic transformation can be performed in three steps: First, a stress free body in the initial configuration $\Omega_0$ is deformed to the current configuration $\Omega$ by loading or an imaginary mapping, and then the deformation tensor $\mathbf{F}$ is obtained. Second, the density $\rho$ and instantaneous elastic tensor $\mathbf{C}$ in the current configuration are calculated using Eq. (\ref{eq2b}). Third, the wave equation will be solved on the current configuration with the derived local material parameters. The numerical simulations are all conducted using PDE (Partial Differential Equation) and Nonlinear Mechanics module in the commercial software COMSOL Multiphysics.
\begin{figure}
	\includegraphics[width=8.5cm]{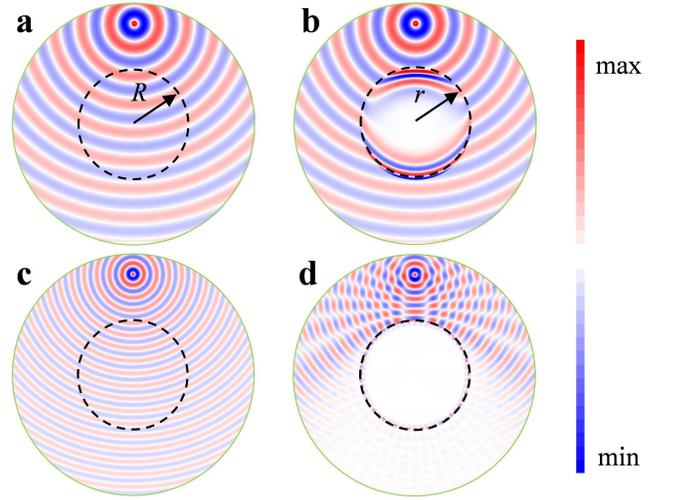}
	\caption{\label{fig1}Wave propagation on the stress-free and pre-deformed semi-linear materials. (a), (b) Divergence fields for the stress-free and pre-deformed cases with longitudinal wave excitation. (c),(d) Curl fields in the stress-free and pre-deformed cases with shear wave excitation.}
\end{figure}

\begin{figure*}
	\includegraphics[width=14cm]{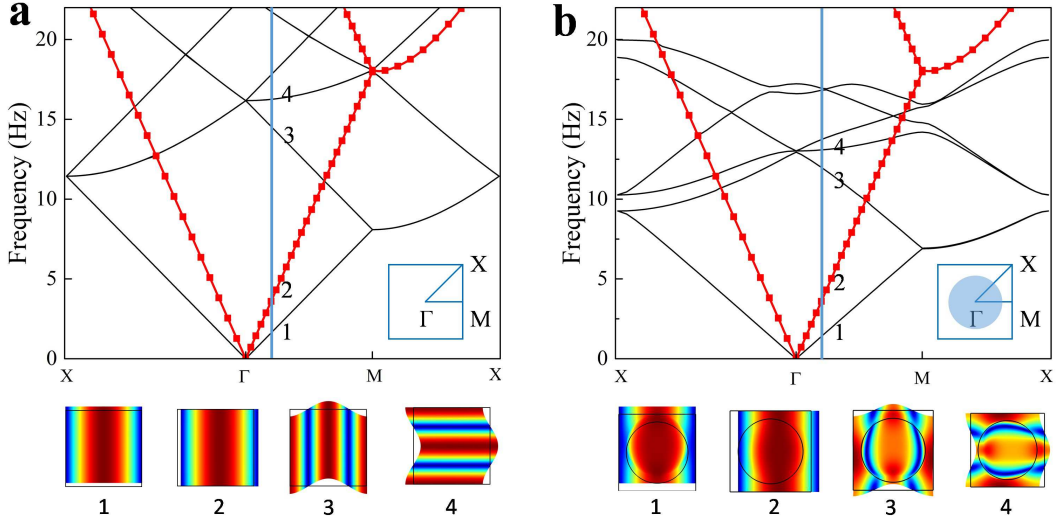}
	\caption{\label{fig2}Band structure of the stress-free and pre-deformed semi-linear materials. (a) Stress free semi-linear material case and the first four modes correspondent to a wave vector  $\mathbf{k}=\boldsymbol{\Gamma}\mathbf{X}/6$. (b) Semi-linear material with pre-deformed circular region and the first four mode correspondent to a wave vector  $\mathbf{k}=\boldsymbol{\Gamma}\mathbf{X}/6$.}
\end{figure*}
We simulate the wave propagation in a pre-deformed semi-linear material that can magnify the wavelength. As shown in FIG. \ref{fig1}(a), the initial configuration consists of two regions, an inner circular region ($R<b=11$ m), where $b$ is the radius of the outer boundary (black dashed line) for the pre-deformed region, and an outer annular region ($b<R<24$ m). Perfectly Matched Layers adjacent to the outer circular boundary is employed to mimic an infinite large space without reflection. With the mapping $r=R^4/b^3$ for the inner circular region ($R<b$), this region is transformed to another circular region with the same radius b as shown in FIG. \ref{fig1}(b). The radial mapping ensures the symmetric deformation gradient $\mathbf{F}$ and therefore the condition of transformation method for longitudinal wave holds. A tiny circle placed at $r=20$ m can excite longitudinal or shear waves ($f=6.5$ Hz) by expansion or rotation motions, respectively. Material parameters of the semi-linear material are $\lambda_0=770$ Pa, $\mu_0=260$ Pa, $\rho_0=1$ kg/m$^3$. In the simulation, the longitudinal and shear wave excitations are modeled through expanding and rotating boundary of the small circle in the domain. Figures \ref{fig1}(a) and \ref{fig1}(b) show divergence fields for the stress free and pre-deformed cases with the longitudinal wave excitation, respectively. Exactly the same fields are observed outside the pre-deformed region as predicted theoretically. Further, the divergence fields in the pre-deformed region ($r<b$) can be perfectly mapped from the fields in the initial configuration with $\nabla_i u_i=F^{-1}_{kL}\nabla_K v_L$. While, the curl fields for the stress free and pre-deformed cases (FIG. \ref{fig1}(c),(d)) with the shear wave excitation are completely different, and significant scattering caused by the pre-deformed region is observed. This interesting phenomenon indicates that the pre-deformation region is invisible for longitudinal waves, but is a strong scatter for shear waves.

Transformation method for longitudinal waves with a pre-deformed semi-linear material is further demonstrated with calculation of dispersion relation. Two square unit cells of the same sizes with a lattice constant $a=1$ m are considered, and the periodic medium are constructed from tiling unit cells along two orthogonal directions. One is composed of a homogeneous stress free semi-linear material, while the other has its inner region ($r<0.35$ m) pre-deformed from the same circular region ($R<0.35$ m) with a mapping function $r=R(R/0.35)^{1/2}$. The Bloch wave frequency can be obtained by solving the eigen-frequency problem on one unit cell with the Bloch wave boundary condition. The dispersion relation is obtained by sweeping the Bloch wave vector along the edges of the first irreducible Brillouin Zone ($\mathbf{X}-\boldsymbol{\Gamma}-\mathbf{M}-\mathbf{X}$). Figure \ref{fig2}(a) is the band structure of the stress-free homogenous unit cell, and FIG. \ref{fig2}(b) presents the result of the partly pre-deformed semi-linear material unit cell. Wave polarizations for different branches can be identified from the modes shown below the band structures. The second branch is longitudinal and the mode shows an overall displacement parallel to the wave vector, while all other three branches represent shear waves. From both band structures, it is shown that the purely longitudinal branches (marked by red dots) are exactly the same for the stress-free and pre-deformed unit cells, while the shear related branches become quite different for both cases. This is expected from the longitudinal wave invariance property of the symmetrically pre-deformed semi-linear material.
\begin{figure}
	\includegraphics[width=8.5cm]{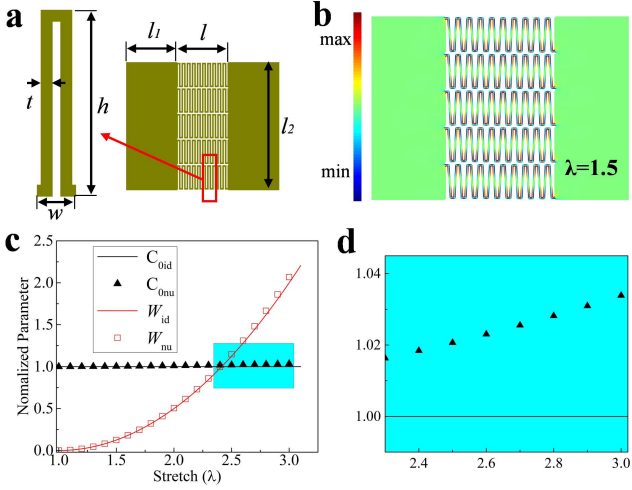}
	\caption{\label{fig3}Spring lattice as a 1D semi-linear material.(a) Unit cell of the U-shaped spring and sandwich spring composite. (b) The deformation field ($\nabla\cdot\mathbf{u}$) of the composite with a finite strain ($\lambda=1.5$). (c) Mechanical property of 10 connected unit cells in (a). (d) Enlarged view of the indicated cyan region in (c) with a strain range ($\lambda=2.3\sim3.0$).}
\end{figure}
\subsection{Prototype of a 1D semi-linear material with spring lattice}
Inspired by the Hooke's law of a perfect spring, a 1D spring lattice is proposed to mimic the semi-linear material. As shown in FIG. \ref{fig3}(a), a unit cell of the proposed spring lattice is composed of connected beams with the following geometry parameters, $t=0.012$ m, $h=0.2$ m and $w=0.04$ m, and all the beams are made of the same material. A sandwiched spring composite (SSC) can be constructed by filling the springs into two separated homogenous materials. A unit cell of the SSC consists of multiple spring unit cells ($5\times10$, $l=10w$) and two layers of the homogeneous materials ($l_1=l$, $l_2=5h$).

The hyper-elastic strain energy function $W(F_{11})$ of the proposed 1D semi-linear material is obtained numerically by solving the stored strain energy in 10 connected unit cells with the applied displacement boundary condition, $u=(F_{11} - 1)10w$, and shows quadratic relation (marked by squares in FIG. \ref{fig3}(c)) with respect to the deformation gradient $F_{11}=\lambda$ for a finite deformation range $F_{11}=1.0\sim3.0$ as compared to the ideal semi-linear case (red line). Two important points should be noted in the numerical simulation. First, since finite rotations occur in this problem during elongation, the geometric nonlinear effect must be turned on during the numerical simulation. Second, although the overall elongation ratio of the unit cell is no longer small, the local strains in the springs and the homogeneous layers are still quite small (the Green strain is of the order of magnitude of $10^{-3}$ for a elongation ratio $\lambda=3$), so a linear elastic constitutive relation is set for the spring material. Here, the constitutive relation is taken to be the default setting in COMSOL with geometry nonlinearity, i.e., St-Venant-Kirchohoff constitutive relation. In short, the hyper-elastic behavior of the proposed 1D spring lattice comes from the finite rotation but small strain of the integrated microstructure. The numerically computed pull-backed modulus $A_{1111}=\partial^2 W/\partial F_{11}/\partial F_{11}$ (marked by triangles) is nearly constant for the investigated finite strain range. The highlighted region in FIG. \ref{fig3}(c) is shown more clearly in FIG. \ref{fig3}(d), where the modulus changes less than 4 percent from a constant value. The numerical result shows the proposed spring lattice has a nearly constant pull-backed modulus under the elongation $F_{11}$, and therefore can be used with the hyper-elastic transformation method to control longitudinal wave along the elongation direction. As for its impact on the longitudinal wave in the other directions, the hyper-elastic strain energy function should be studied for a general deformation gradient and this is beyond the main scope of this paper. Finally, since the pull-backed modulus is obtained from the static homogenization or under long wave approximation, so in the following analysis, we consider only the low frequency case where the static homogenization condition holds. 
\begin{figure}
\includegraphics[width=8.5cm]{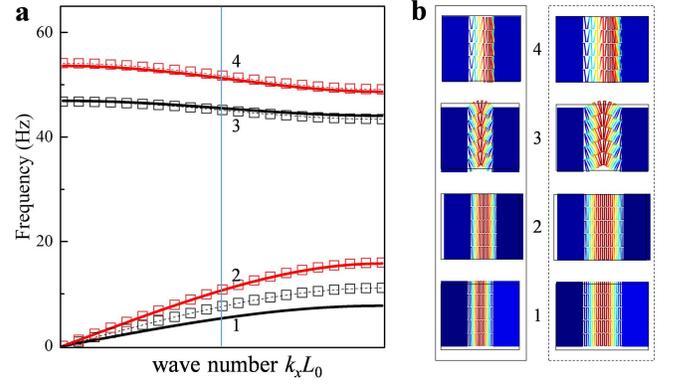}
\caption{\label{fig4}Band structure of the proposed 1D semi-linear material. (a) Band structure of the SSC unit cell without (line) and with (discrete square) pre-deformation. (b) Modes of the first four branches correspondent to a normalized wave number $k_x L_0=\pi/2$, left panel is for the stress-free case and right panel is for the pre-deformed case.
}
\end{figure}

Figure \ref{fig3}(b) shows the deformed state of the SSC with a finite strain $\lambda=1.5$, where the deformation mainly concentrates in the turning points of the springs and the strain in the homogenous material is extremely small, therefore the springs can be regarded as pre-deformed while the layered materials remain in its initial configuration. The dispersion relations for the un-deformed and pre-deformed SSC unit cells (FIG. \ref{fig3}(b)) are shown in FIG. \ref{fig4}(a) together with their corresponding deformation modes in FIG. \ref{fig4}(b). Since the unit cells of the un-deformed and pre-deformed cases have different lengths, the horizontal axis in FIG. \ref{fig4}(a) represents a non-dimensional wave number $k_x L_0$ with $L_0$ being the length of the corresponding unit cell. The true deformation is rather small and has been exaggerated in the figures for clarity. The modes indicate that the second and forth branches are mainly related to the longitudinal waves and therefore these two branches of the pre-deformed case are almost the same as the stress-free case. While the first and third branches are dominated by shear waves as shown by the modes, and are changed during stretching, especially for the first branch. The above results show that the SSC can be used as a robust semi-linear material for controlling longitudinal waves with the hyper-elastic transformation method.

\section{CONCLUSIONS}
The hyper-elastic transformation method with semi-linear materials is revisited. It is proved that semi-linear materials under a general symmetric pre-deformation maintain the form invariance for longitudinal elastic waves, while not for shear waves. This implies that the longitudinal wave control can be made with the hyper-elastic transformation method without the additional constraint on pre-stretches. Wave simulations on a pre-deformed semi-linear material show that, the longitudinal waves follow exactly the designed path by the hyper-elastic transformation method without any scattering, while the shear waves are strongly scattered. The dispersion relation of a semi-linear material, with a symmetric pre-deformation, is shown invariant for the longitudinal wave branches, but not for the shear related branches. These numerical results confirm the theoretical finding. Finally, a 1D semi-linear material realized by a spring lattice is proposed to mimic a semi-linear material. Numerical simulations reveal that the longitudinal wave branches remain intact during stretching. The demonstrated property of the semi-linear material may find applications where longitudinal and shear waves are expected to be controlled differently with external deformation.

\begin{acknowledgments}
This work was supported by National Natural Science Foundation of China (Grant Nos. 11472044, 11521062, 11632003).
\end{acknowledgments}

\appendix*
\section{\label{appen}DERIVATIVE OF STRAIN ENERGY FUNCTION}

Following the polar decomposition of a deformation tensor $\mathbf{F}=\mathbf{R}\mathbf{U}$, we have $\mathbf{F}^{\mathbf{T}}\mathbf{F}=\mathbf{U}^2$ or in component form $U_{IK}U_{KJ}=F_{kI}F_{kJ}$. Taking the derivative of both sides with respect to $F_{lN}$ leads to,
\begin{equation}
\label{eqA1}
\frac{\partial U_{IK}}{\partial F_{lN}}U_{KJ}+\frac{\partial U_{KJ}}{\partial F_{lN}}U_{IK}=F_{lI}\delta_{Nj}+F_{lJ}\delta_{Nl} 
\end{equation}
From det($\mathbf{U}$)=det($\mathbf{F}$)>0, one can deduce that $\mathbf{U}$ is reversible. Multiplying both sides with $\delta_{IJ}$ and $U^{-1}_{IJ}$ gives, respectively,
\begin{equation}
\label{eqA2}
\frac{\partial U_{IK}}{\partial F_{lN}}U_{KI}=F_{lN}
,\quad \frac{\partial U_{KK}}{\partial F_{lN}}=F_{lI}U^{-1}_{IN}=R_{lN}
\end{equation}
Further differentiating Eq. (\ref{eqA1}) with respect to $F_{rS}$ and multiplying both sides with $U^{-1}_{IJ}$ gives,
\begin{equation}
\label{eqA3}
\frac{\partial R_{lN}}{\partial F_{rS}}+\frac{\partial U_{IK}}{\partial F_{lN}}\frac{\partial U_{KJ}}{\partial F_{rS}}U^{-1}_{IJ}=\delta_{lr}U^{-1}_{SN}
\end{equation}
If $\mathbf{F}$ becomes an identity matrix, Eq. (\ref{eqA3}) can be simplified as,
\begin{equation}
\label{eqA4}
\frac{\partial R_{lN}}{\partial F_{rS}}|_{\mathbf{F}=\mathbf{I}}=\frac{1}{2}(\delta_{lr}\delta_{SN}-\delta_{lS}\delta_{rN})
\end{equation}
With above formulas in Eq. (\ref{eqA2}), the second derivative of the semi-linear strain energy function can be derived,
\begin{widetext}
\begin{equation}
\label{eqA5}
\begin{split}
A_{MjNl}&=\frac{\partial^2 W}{\partial F_{jM}\partial F_{lN}}=\frac{\partial}{\partial F_{jM}}\frac{\partial}{\partial F_{lN}}(\frac{\lambda_0}{2}(U_{KK}-\delta_{KK})^2+\mu_0(U_{KL}U_{KL}-2U_{KK}))\\
&=\frac{\partial}{\partial F_{jM}}(\lambda_0(U_{KK}-\delta_{KK})\frac{\partial U_{KK}}{\partial F_{lN}}+2\mu_0 U_{KL}\frac{\partial U_{KL}}{\partial F_{lN}}-2\mu_0\frac{\partial U_{KK}}{\partial F_{lN}})\\
&=\frac{\partial}{\partial F_{jM}}((\lambda_0U_{KK}-\lambda_0\delta_{KK}-2\mu_0)R_{lN}+2\mu_0F_{lN})\\
&=\lambda_0R_{jM}R_{lN}+2\mu_0\delta_{lj}\delta_{MN}+(\lambda_0U_{KK}-\lambda_0\delta_{KK}-2\mu_0)\frac{\partial R_{lN}}{\partial F_{jM}} 
\end{split}
\end{equation}
\end{widetext}
If the deformation tensor becomes symmetric $\mathbf{F}=\mathbf{F}^\mathbf{T}$, we can simplify the pull-backed modulus Eq. (\ref{eqA5}) as,
\begin{equation}
\label{eqA6}
\begin{split}
A_{MjNl}|_{\mathbf{F}=\mathbf{F}^\mathbf{T}}=&\lambda_0\delta_{jM}\delta_{lN}+2\mu_0\delta_{ij}\delta_{MN}\\
+&(\lambda_0(U_{KK}-\delta_{KK})-2\mu_0)\frac{\partial R_{lN}}{\partial F_{jM}}
\end{split}
\end{equation}
Notice that, $\partial R_{lN}/\partial F_{jM}=\partial R_{jM}/\partial F_{lN}$ is skew-symmetric with respect to index $j$ and $M$ when $\mathbf{F}$ is symmetric. It can be easily proved from this mathematical identity $\mathbf{R}\mathbf{R}^{\mathbf{T}}=\mathbf{I}$ by differentiation,
\begin{subequations}
\begin{equation}
\label{eqA7}
\frac{\partial R_{iK}}{\partial F_{lN}}R_{jK}+R_{iK}\frac{\partial R_{jK}}{\partial F_{lN}}=0
\end{equation}
\begin{equation}
\label{eqA8}
\frac{\partial R_{iJ}}{\partial F_{lN}}|_{\mathbf{F}=\mathbf{F}^{\mathbf{T}}}+\frac{\partial R_{jI}}{\partial F_{lN}}|_{\mathbf{F}=\mathbf{F}^{\mathbf{T}}}=0
\end{equation}
\end{subequations}
Further if $\mathbf{F}$ becomes an identity matrix, the pull-backed modulus becomes the conventional isotropic one,
\begin{equation}
\label{eqA9}
\begin{split}
A_{MjNl}|_{\mathbf{F}=\mathbf{F}^{\mathbf{T}}}=&\lambda_0\delta_{jM}\delta_{lN}+2\mu_0\delta_{lj}\delta_{MN}-2\mu_0\frac{\partial R_{lN}}{\partial F_{jM}}|_{\mathbf{F}=\mathbf{I}}\\
=&\lambda_0\delta_{jM}\delta_{lN}+\mu_0\delta_{lj}\delta_{MN}+\mu_0\delta_{lM}\delta_{jN}
\end{split}
\end{equation}

%
\end{document}